# Tunable two-dimensional superconductivity and spin-orbit coupling at the EuO/KTaO$_3$(110) interface

Xiangyu Hua,[1] Fanbao Meng,[1] Zongyao Huang,[1] Zhaohang Li,[1] Shuai Wang,[3] Binghui Ge,[3] Ziji Xiang,[1,*] and Xianhui Chen[1,2,*]

[1]*Department of Physics, University of Science and Technology of China, and Key Laboratory of Strongly Coupled Quantum Matter Physics, Chinese Academy of Sciences, Hefei 230026, China*

[2]*Collaborative Innovation Center of Advanced Microstructures, Nanjing University, Nanjing 210093, China*

[3]*Information Materials and Intelligent Sensing Laboratory of Anhui Province, Institutes of Physical Science and Information Technology, Anhui University, Hefei, 230601, China*

Unconventional quantum states, most notably the two-dimensional (2D) superconductivity, have been realized at the interfaces of oxide heterostructures where they can be effectively tuned by the gate voltage ($V_G$). Here we report that the interface between high-quality EuO (111) thin film and KTaO$_3$ (KTO) (110) substrate shows superconductivity with onset transition temperature $T_c^{onset}$ = 1.35 K. The 2D nature of superconductivity is verified by the large anisotropy of the upper critical field and the characteristics of a Berezinskii-Kosterlitz-Thouless transition. By applying $V_G$, $T_c^{onset}$ can be tuned from ~ 1 to 1.7 K; such an enhancement can be possibly associated with a boosted spin-orbit energy $\varepsilon_{so} = \hbar/\tau_{so}$, where $\tau_{so}$ is the spin-orbit relaxation time. Further analysis of $\tau_{so}$ based on the upper critical field ($H_{c2}$) and magnetoconductance reveals complex nature of spin-orbit coupling (SOC) at the EuO/KTO(110) interface with different mechanisms dominate the influence of SOC effects for the superconductivity and the magnetotransport in the normal state. Our results demonstrate that the SOC should be considered as an important factor determining the 2D superconductivity at oxide interfaces.

## INTRODUCTION

Oxide heterojunctions are ideal platforms for exploring a remarkable variety of emergent phenomena.[1] In particular, the superconducting electron gases residing at the interface between two insulators have attracted considerable attention.[2-9] Previous studies have revealed that the conductive interfaces in the SrTiO$_3$ (STO)-based heterojunctions exhibit a surprisingly enriched cascade of unusual properties including the coexistence of superconductivity and ferromagnetism,[10-12] and pseudogap-like behaviour.[13] The more recently discovered KTaO$_3$ (KTO)-based heterojunctions also host two-dimensional (2D) superconductivity[8,9] and anomalous Hall effect.[14] Most interestingly, application of electric fields can effectively modulate the strength of spin-orbit coupling (SOC),[15-17] the carrier density,[18,19] and the disorder level,[20] consequently it can control the $T_c$ and even lead to a superconducting-to-insulator quantum phase transition.[18-20] Despite that KTO and STO are isostructural compounds with similar band structures,[21] the two systems are distinct from each other in many aspects. The conductive electrons in STO and KTO are contributed by the Ti 3d and Ta 5d bands, respectively,[22] thus the KTO-based superconducting interface (SI) should have a stronger SOC. Furthermore, whilst superconductivity has been realized at the (001)-, (110)- and (111)-oriented STO-based interfaces with comparable $T_c$,[2,3,7] the KTO (001)-based interface is non-superconducting,[14,23] whereas the KTO (110) and KTO (111)-based interfaces show superconductivity with $T_c$ around 1 and 2 K, respectively.[8,9]

Several recent experimental observations, e.g., peculiar stripe phase in the normal state (although needs further confirmation)[8,24] and unusual doping dependence of the upper critical field ($H_{c2}$),[25] imply possible unconventional superconductivity at the KTO-based interface. Further evidence for the novelty of the superconductivity can stem from its responses to the external electric fields. It has been argued that the back gate voltage applied to the LaAlO$_3$/KTO(111) (LAO/KTO) interface predominantly controls the effective disorder (and consequently the electron scattering rate) rather than the carrier density.[20] At the EuO/KTO(111) interface, both the $T_c$ and $H_{c2}$ are reported to be sensitive to the carrier density in the gating process.[25] On the other hand, investigations of the electric-field-control SOC, which may be suggestive of unconventional superconducting pairing mechanisms,[17] are still lacking for the KTO-based devices.

Under certain circumstances, the SOC effect can play an essential role in determining the physics in oxide heterostructures (most notably at LAO/STO interface close to the Lifshitz point)[26], due to the inversion symmetry breaking at the interface.[15-17] For these interfaces, the behaviours of both the normal and the superconducting states are to be understood only with the SOC taking into account. For instance, the magnetotransport properties of the normal state usually reflects the influence of SOC in terms of the weak localization or weak antilocalization effects,[27,28] whereas the unusually high $H_{c2}$ and potential unconventional pairing in the superconducting state can be triggered by the complex contributions of SOC.[17,29]

Moreover, it has been shown that the SOC at the STO-based SIs can be effectively manipulated by the application of electric fields, leading to intricate evolution of physical properties as revealed experimentally.[15-17] In this sense, it is important to explore the role of SOC at the KTO-based SIs and its influence on the 2D superconductivity.[8,25]

This work reports the growth of high-quality EuO (111) thin films on KTO (110) substrates. The interface between them is proved to host 2D superconductivity. Most intriguingly, we find a large tunability of $T_c$ and SOC at the EuO/KTO(110) interface by applying an electric field across KTO substrates. Based on these observations, we propose that in addition to carrier density and effective disorder, the SOC strength has a significant impact on $T_c$. This may be linked to the unconventional nature of the superconductivity at the KTO-based interface.

**RESULTS AND DISCUSSION**
**Characterization of the EuO/KTO(110) heterostructures**

High-quality EuO (111) thin films were grown on (110)-oriented KTO substrates using a molecular beam epitaxy (MBE) system (see Methods for details). Bulk EuO crystallizes in a cubic structure with lattice constant $a$ = 5.145 Å. Stoichiometric EuO is an insulator with a band-gap of 1.12 eV at room temperature.[30] Figure 1a shows a schematic illustration for our EuO/KTO(110) heterostructure. To confirm the quality of films, we performed the scanning transmission electron microscopy (STEM) measurements. Figure 1b shows a cross-section with KTO [001] orientation in plane (another direction along KTO [1$\bar{1}$0] is shown in Fig. S1a). Due to a certain lattice mismatch (~ 5%) between EuO (111) and KTO (110) surfaces, the epitaxial EuO at the interface is distorted within a thickness of approximately 2 atomic layers (red square in Fig. S2); single crystallinity is recovered beyond this region. Atomic-scale energy-dispersive x-ray spectroscopy (EDS) shows a relatively clear interface in which the diffusion of Eu exists in the superficial layer of KTO (the white dotted region in Fig. S2). Electron energy loss spectroscopy (EELS) peaks of Eu also suggest that the Eu doping in KTO persists up to approximately 3 atomic layers crossing the interface (Fig. S3). $\theta$-$2\theta$ X-ray diffraction (XRD) pattern confirms that our samples have good single crystallinity (Fig. 1c). A fit using the angle of the Laue oscillation peaks yields the film thickness of about 7 nm (Fig. S1b). The films also exhibit good surface flatness with a root-mean-squared roughness around 0.343 nm (Fig. S4).

The transport properties were measured using the Van der Pauw method (inset of Fig. 1d). The samples are metallic in the whole temperature ($T$) range as shown in Fig. 1d, indicating the formation of electron gases at the interfaces. Both samples undergo a superconducting transition at low temperatures. For sample #1, $T_c^{onset}$ is 1.35 K, and the zero resistance is observed at $T_c^{zero}$ = 1.06 K (Fig. 1e). The magnetic-field ($H$)-dependent Hall resistance $R_{Hall}$ measured at $T$ = 2 K confirms that the charge carriers are electron-type for both samples (inset of Fig. 1e). In Fig 1f we plot the 2D Hall carrier density $n_s$ and the Hall mobility $\mu$ (extracted from the Hall and sheet resistance data) versus temperature. For sample #1 (#2), $n_s$ is 8.6 (9.0)×10$^{13}$ cm$^{-2}$ and $\mu$ is 86 (128) cm$^2$/V s at 2 K. Compared to LAO/KTO(110) interface,[9] both $n_s$ and $T_c$ in our samples are higher, consistent with the results for the (111)-oriented devices.[8] During the growth process, Eu atoms have a strong capability to uptake oxygen from the surface layer of the KTO; this effect may cause the higher $n_s$ (and consequently the enhanced $T_c$) in the EuO/KTO heterostructures.

**2D superconductivity**

We measured the $T$-dependent 2D sheet resistance $R_{sheet}$ under magnetic fields applied perpendicular and parallel to the interface to investigate the nature of this interfacial superconductivity. As shown in Figs. 2a and 2b, the superconductivity is remarkably suppressed by a magnetic field of ~0.4 T and ~6 T applied perpendicular and parallel to the interface, respectively. Such strong anisotropy indicates the 2D nature of superconductivity. To further verify this, we fit our data to the Ginzburg-Landau theory for a 2D superconductor:[31]

$\mu_0 H_{c2}^{//c}(T) = [\Phi_0/2\pi\xi_{GL}^2(0)][1-(T/T_c)]$,
$\mu_0 H_{c2}^{//ab}(T) = [\Phi_0\sqrt{12}/2\pi\xi_{GL}(0)d_{sc}][1-(T/T_c)]^{1/2}$,   (1)

where $\xi_{GL}$ is the Ginzburg-Landau coherence length, and $\Phi_0$ is the flux quantum, and $d_{sc}$ is the superconducting layer thickness. The $T/T_c$ dependence of upper critical fields for $H//ab$ ($\mu_0 H_{c2}^{//ab}$) and $H//c$ ($\mu_0 H_{c2}^{//c}$) (Fig. 2c) are determined from the $R_{sheet}$-$T$ curves shown in Figs. 2a and 2b, respectively. The fits to Eq. (1) yield the zero temperature limit values $\mu_0 H_{c2}^{//c}(0)$ = 0.45 T and $\mu_0 H_{c2}^{//ab}(0)$ = 6.65 T, corresponding to an anisotropic ratio of ~ 15; meanwhile, the fits also give $\xi_{GL}(0)$ = 27.03 nm and $d_{sc}$ = 6.35 nm. A similar analysis for sample #2 were shown in Fig. S5. Since the growth conditions of #1 and #2 are different, $\mu_0 H_{c2}$ and $d_{sc}$ of these two samples also differ slightly. Nonetheless, $d_{sc}$ is smaller than $\xi_{GL}$ in both samples, confirming the 2D nature of the superconductivity at EuO/KTO(110) interface. Besides, $d_{sc}$ is much larger than the diffusion depth (~0.4 nm) of Eu atoms (Fig. S3). The mean free path of the conducting electrons can be estimated as $l_{mfp} = (h/e^2)(1/k_F R_{sheet})$ in a single-band model ($k_F = \sqrt{2\pi n_s}$ is the Fermi wave number, $h$ is the Planck constant, $e$ is the elementary charge).[32] Using the measured $R_{sheet}$(2 K) and $n_s$(2 K), we estimated the $l_{mfp}$ of sample #1 to be 59.3 nm, which is larger than the $\xi_{GL}$. We note that though our EuO/KTO(110) interfaces are much cleaner compared to the LAO/KTO(110) and EuO/KTO(111) interfaces wherein $l_{mfp} < \xi_{GL}$[8,9], these SIs are still in the dirty limit (Methods).

With solid evidence for 2D superconductivity, we further examine the expected behaviours of the Berezinskii-Kosterlitz-Thouless (BKT) transition in our devices.[33] The BKT transition, a transition from unpaired vortexes and anti-vortexes to bound vortex-antivortex pairs, can result in a $V \propto I^\alpha$ power-law dependence and can be characterized by a transition temperature $T_{BKT}$ where $\alpha(T_{BKT}) = 3$.[34] To reveal such characteristics, we measure the current-dependent voltage (I-V curves). The data for sample #1 is displayed in Fig. 3a (see Fig. S6a for sample #2). By fitting of the I-V curve in the nonlinear range (Figs. 3b and 3c), we attain an exponent α approaching 3 at $T_{BKT}$ = 1.01 K. Apart from the I-V method, the $T_{BKT}$ can also be estimated from the formula $R_{sheet}(T) = R_0 \exp[-b(T/T_{BKT}-1)^{-1/2}]$, where $R_0$ and $b$ are material parameters. Application of such fit to the measured $R_{sheet}(T)$ yields $T_{BKT}$ = 1.17 K (inset of Fig. 3c). $T_{BKT}$ obtained from these two approaches appears to be close to $T_c^{zero}$, again pointing towards the 2D nature of the superconductivity.

For a 2D weak coupling BCS superconductor, the parallel critical field can be determined by the Chandrasekhar-Clogston limit (Pauli paramagnetic limit)[35,36]: $\mu_0 H_P \approx 1.76 k_B T_c / \sqrt{2} \mu_B$, where $k_B$ and $\mu_B$ are the Boltzmann's constant and Bohr magneton, respectively. Taking $T_c = T_{BKT}$ = 1.01 K, we have $\mu_0 H_P$ = 1.874 T, which reaches only 28% of $\mu_0 H_{c2}^{//ab}(0)$ (blue dashed line in Fig. 2c). Several factors can enhance the Pauli limit appreciably, such as strong-coupling superconductivity and many-body effects.[37,38] In our samples, the most likely reason for the large $\mu_0 H_c^{//ab}$ could be the strong SOC originating from the inversion symmetry breaking at the interface and the relatively heavy tantalum ions,[29,39] which can be verified by the electric-field manipulation that we will discuss in the following section.

**Electric-field control of superconductivity**

To explore the effect of electric fields, our samples were made into 20 × 100 μm² Hall bar devices (inset of Fig. 4a). A schematic diagram of the device (Fig. S7) and the manufacturing process is presented in Methods. The superconductivity can be successfully tuned by applying a gate voltage ($V_G$) across KTO (Fig. 4a). As $V_G$ varies from -180 to 150 V, $T_c^{onset}$ ($T_c^{mid}$) increases from 0.96 (0.89) to 1.69 (1.54) K, respectively, highlighting an enhancement exceeding 70% (Fig. 4b). Meanwhile, at $T = 3.8$ K, $n_s$ shows a moderate change, achieving the maximum value at $V_G = -80$ V, whereas $\mu$ varies monotonically from 73 to 211 cm$^2$/V s over the entire gating voltage range (Fig. 4c). As $R_{Hall}(H)$ maintains linearity at all gate voltages (Fig. S8b), the mean-free path $l_{mfp}$ and the effective disorder $k_F l_{mfp}$ at different voltages can be estimated using the 3.8 K data of $R_{sheet}$ and $n_s$. Over the ramping range of $V_G$ from 150 to -180 V, $k_F l_{mfp}$ evolves from 75 to 27 and $l_{mfp}$ varies from 32.2 to 11.4 nm (Fig. S9). Such variation of $k_F l_{mfp}$ and $l_{mfp}$ is much smaller than that of LAO/KTO(111) interface.[20] For the latter, it is shown that the tuning effect of electric fields highly depends on the mobility (disorder level) of the sample.[20] As the mobility is relatively high in our single-crystalline heterostuctures with cleaner interfaces, the electric-field controlling of the effective disorder scattering is less efficient compared to that reported in ref. [20].

The limited ability of the electric fields in controlling the carrier density and disorder suggests that there should be an alternative origin for the continuous increase of $T_c$, especially for $V_G < -80$ V where $n_s$ and $T_c$ show anticorrelated behaviour. In LAO/STO systems, the spin-orbit coupling, whose strength directly affects $\mu_0 H_{c2}^{//ab}$, is believed to contribute to stabilizing the 2D superconductivity.[29,39,40] To examine the validity of such scenario in our devices, we trace the evolution of $\mu_0 H_{c2}^{//c}$, $\mu_0 H_{c2}^{//ab}$, and spin-orbit energy ($\varepsilon_{so}$) during the gating process. As shown in Fig. 4d, We plot the comparison between $\mu_0 H_{c2}^{//c}$ (inset of Fig. 4d), $\mu_0 H_P$, and $\mu_0 H_{c2}^{//ab}$ (Fig. 4d) (at $T = 0.1$ K, calculated using the methods shown in Figs. S10 and S11) upon varying $V_G$. The much higher $\mu_0 H_{c2}^{//ab}$ relative to $\mu_0 H_P$ (Fig. 4d) is likely to be caused by a strong spin-orbit interaction and/or $V_G$-dependent electron wave function widths at the interface (which can be understood as the superconducting layer thickness $d_{sc}$): in an effective model for 2D superconductor with strong SOC, we have[17,41]:

$$\mu_0 H_{c2}^{//ab}(T=0) = \sqrt{\frac{1.76\hbar k_B T_c}{[3\mu_B^2 \tau_{so} + D(d_{sc}e)^2/3]}} \qquad (2)$$

where $\tau_{so}$ is the spin-orbit relaxation time ($\varepsilon_{so} = \hbar/\tau_{so}$), $D$ is the diffusion constant obtained from the slope of the out-of-plane upper critical field: $[-d(\mu_0 H_{c2}^{//c})/dT]_{T=T_c} = 4k_B/\pi D e$ (we use the data in Fig. S5c to fit $D$ because sample #2 has similar $\mu_0 H_{c2}$ with the gating sample)[41]. The fits of $\mu_0 H_{c2}$ using eq. (1) allow us to determine the evolution of $\xi_{GL}$ and $d_{sc}$ upon the continuous changing of $V_G$ from 150 to -180 V (Fig.

4e). As the $V_G$ varies from 150 to -180 V, $\xi_{GL}$ ($d_{sc}$) decreases from 32 (14) to 22 (2.7) nm (an illustration of $V_G$-dependent $d_{sc}$ is shown in Fig. S12), $\xi_{GL}$ is much larger than $d_{sc}$ at all $V_G$, especially for $V_G < 0$, confirming the 2D nature of the superconductivity at EuO/KTO (110) interface in the gating process. Taking $T_c = T_c^{mid}$, we also obtain the $V_G$-dependent $\varepsilon_{so}$ as plotted in Fig. 4d. With $V_G$ ramping from 150 to -180 V, $\varepsilon_{so}$ increases from 1.2 to 21.1 meV with a large ratio of ~17.6, revealing a strong tunability of the SOC at the EuO/KTO(110) interface. The plot of $T_c$ versus $\varepsilon_{so}$ (Fig. 4f) indicates that $T_c$ is predominantly controlled by different factors in two ranges of $V_G$ separated by $V_G = -80$ V where $T_c$ ($\varepsilon_{so}$) exhibits a clear kink. From $V_G = 150$ to -80 V, $T_c$ rises rapidly with decreasing $V_G$. Considering the increasing $n_s$ with decreasing $V_G$ in this range, the increase of $T_c$ is predominantly driven by the variation of carrier density. By contrast, as $V_G$ is swept from -80 towards -180 V, $T_c$ increases linearly with $\varepsilon_{so}$ (Fig. 4f), whereas $n_s$ gradually decreases. Therefore, we conclude that the increase of $T_c$ is directly related to the enhancement of spin-orbit scattering in the range of $V_G$ from -80 to -180 V (Fig. 4b).

The present work is the first report of the electric-field-controlled SOC at the KTO-based SI. Here we briefly compare it with that in the STO-based SI. Distinct from the barely tunable SOC at the LAO/STO(110) interface,[42] the tunability of SOC in our samples is considerably large. Nonetheless, a relationship between $T_c$ and $\varepsilon_{so}$ similar to that is displayed in Fig. 4d has been observed in LAO/STO(100) SI (see Figs. S13a-c for our analysis based on the published data).[15,16] We also noticed that, for the LAO/STO(111) SI, a dome-shaped $V_G$-dependence of both $T_c$ and $\varepsilon_{so}$ has been established,[17] with a roughly linear relationship between these two (Fig. S13d); such a close link between $T_c$ and $\varepsilon_{so}$ is proposed to hint at potential unconventional superconductivity pairing mechanism.[17,43,44] In our case, the unconventional linear relationship between the electric-field-control SOC and superconductivity implies the possibility of unconventional pairings at the KTO-based SIs, which definitely deserves further studies.

**SOC relaxation time in magnetotransport**

The strength of SOC can also be evaluated from the normal-state perpendicular magnetoresistance. Here, we measured the magnetoresistance at $T = 3.8$ K under different $V_G$ (Fig. S14). In the diffusive regime, the field-dependent quantum correction to conductivity $\Delta\sigma(H)$ can be described by the Maekawa-Fukuyama (MF) model[17,45]:

$$\frac{\Delta\sigma(H)}{\sigma_0} = \Psi\left(\frac{H}{H_i+H_{so}}\right) + \frac{1}{2\sqrt{1-\gamma^2}}\Psi\left(\frac{H}{H_i+H_{so}(1+\sqrt{1-\gamma^2})}\right) - \frac{1}{2\sqrt{1-\gamma^2}}\Psi\left(\frac{H}{H_i+H_{so}(1-\sqrt{1-\gamma^2})}\right) - \frac{AH^2}{1+CH^2}$$

(3)

here $\sigma_0 = 2e^2/h$ is the quantum conductance, $\Psi(x) = \ln(x) + \Psi[1/2 + (1/x)]$ [ $\Psi(x)$ is a digamma function and $\gamma = g\mu_B H/4eD_{tr}H_{so}$ ], $H_i$ and $H_{so}$ are the inelastic and spin-orbit effective fields, respectively. The last term including parameters $A$ and $C$ is a Kohler term originated from the classical orbital magnetoresistance. Combining the 2D nature of the superconductivity in the gating process, the diffusion coefficient $D_{tr}$ can be expressed as: $D_{tr} = v_F^2\tau/2$ ( $v_F = \hbar\sqrt{2\pi n_s}/m^*$ is the Fermi velocity, where $m^*$ is the effective electron mass). The relaxation time $\tau$ for elastic scattering can be extracted from $R_{sheet}$ based on the Drude model: $\tau = m^*/e^2 n_s R_{sheet}$. By applying the fits to the MF model (Eq.3) to the magnetoconductance (Fig. 5a), we obtained the parameters $H_{i,so}$ (Fig. 5b) and $A,C$ (Fig. S15) at different voltages by assuming $g = 2$ and $m^* = m_e$[46] (the variation of $m^*$ within the reasonable range (~0.5-1.0 $m_e$) does not change the qualitative conclusions, see Methods). $H_{so}$ increases with $V_G$ decreasing, in agreement with the $V_G$ dependence of $\varepsilon_{so}$ shown in Fig. 4d and verifies the enhancement of SOC at large negative $V_G$.

The evolution of spin-orbit relaxation time $\tau_{so}$ and the inelastic relaxation time $\tau_i$ can be further derived from the effective fields: $H_{so,i} = \hbar/4eD_{tr}\tau_{so,i}$. In Fig. 5c we plot all three relaxation times $\tau_i$, $\tau_{so}$ and $\tau$ against $V_G$. $\tau_{so}$ is the smallest among them, means that the spin-orbit scattering in the KTO-based SI is strong and dominates the decoherence process. More intriguingly, as shown in Fig. 5d, we have $\tau_{so}^{-1} \propto \tau$; this is consistent with the expectation for the D'yakonov-Perel' (DP) mechanism of spin relaxation.[47] The DP scenario describes the spin precession around the spin-orbit field between scatterings that leads to the spin dephasing; such mechanism is consistent with Rashba-type SOC at the interface.[47] However, we mention that if we plot the $\tau_{so}$ extracted from $\varepsilon_{so}$ (Fig. 4d) determined from $H_{c2}$ against $\tau$ (Fig. 5d), it shows $\tau_{so} \propto \tau$, i.e., $\tau_{so}$ obtained from $H_{c2}$ and the magnetoconductance exhibit distinct behaviours. The relationship of $\tau_{so} \propto \tau$ corresponds to the Elliott-Yafet (EY) mechanism[48,49] describing spin-flip scatterings. Hence, the spin-orbit scattering that affect the pair-breaking effect of the Zeeman field and that contributes to the quantum correction of charge transport in the normal state are assigned to the EY and DP mechanisms, respectively. Possible explanation for this discrepancy is that Cooper pair formation and the normal-state electrical transport are dominated by electrons occupying different conduction channels or subbands at the interface. Similar behaviour has also been observed for the LAO-STO system.[41] More exotic probabilities, such as SOC-enhanced spin susceptibility in the superconducting state (which naturally enhanced the Pauli limit critical field) or unconventional superconducting pairings,[50-52] are to be verified by future investigations.

**SUMMARY AND OUTLOOK**

To conclude, high-quality single-crystalline EuO (111) thin films have been grown on KTO (110) substrates. The large anisotropy of $H_{c2}$ and the characteristics of a BKT transition show that the interface between them is a 2D superconductor. The remarkable response of $T_c$ to the applied $V_G$ is proved to be predominantly linked to the high tunability of SOC strength under external electric fields. $\tau_{so}$ obtained from $H_{c2}$ and the magnetoconductance manifests the typical behaviours expected for the EY and DP spin-relaxation mechanisms, respectively, implying the complexity of the SOC effects at the EuO-KTO SIs. Our results demonstrate that the SOC should be considered as an important factor controlling the 2D superconductivity and might lead to unconventional superconductivity at the KTO-based interface. Further theoretical investigations are needed to elucidate such unusual interplay between the electric-field-control SOC and superconductivity.

**METHODS**

**Growth of EuO/KTO(110) heterostructures and device fabrication.**

EuO (111) thin films were grown on (110)-oriented KTO single crystal using a molecular beam epitaxy system with a base pressure of $4 \times 10^{-10}$ mbar. The samples size are $5 \times 5$ mm$^2$. Before growth, the KTO substrates were pre-annealed at 600 ℃ for 1 hour and then cooled down to growth temperature. The deposition rate of Eu was 0.2 Å/s, calibrated by a quartz-crystal monitor. The depositions were performed at 400 ℃. The oxygen pressures during the growth of the two samples are $1.9 \times 10^{-9}$ mbar for #1, $2.0 \times 10^{-9}$ mbar for #2. After growth, the samples were cooled down to room temperature with no oxygen supply. A 3~4-nm-thick germanium were prepared to protect sample from further oxidation when exposed to air.

The hall bar devices were prepared using standard optical lithography and Argon etching techniques. The etching thickness is 40 nm which is much larger than the thickness of EuO films.

**Scanning transmission electron microscopy (STEM) and transport measurements.**

The slices for STEM were prepared from selected areas using Carl Zeiss Crossbeam 550L and the high angle annular dark field scanning transmission electron microscopy (HADDF STEM) were obtained using a probe Cs-corrected JEOL-ARM200F NEOARM.

The transport measurements were carried out in a 3He cryostat (HelioxVT, Oxford Instruments) and a commercial Quantum Design PPMS with a dilution refrigerator insert.

**Verification of the dirty-limit scenario.**

For a weak coupling BCS superconductor, the BCS coherence length $\xi_{BCS} = (\hbar v_F)/(\pi 1.76 k_B T_c)$. In the negative $V_G$ regime for our sample, $l_{mfp}/\xi_{BCS} = (\pi 1.76 k_B T_c \tau)/\hbar$ ranges from 0.05 to 0.07. Thereby, the condition $l_{mfp} \ll \xi_{BCS}$ for the dirty-limit superconductors is still valid at our SIs. This result validates the application of Eq. (2) in the main text which describes the upper critical field of a 2D superconductor in the dirty limit.

**Effective mass of electrons.**

Due to the relatively low mobility of our samples, our available experimental probes fail to resolve quantum oscillations in magnetoresistance down to 0.1 K (Fig. S16). In the ref. [46], the effective mass of electrons $m^* \approx 0.62 m_e$ under high magnetic fields. To verify the influence of the effective mass of electrons on our experimental conclusions, we assumed $m^* = 0.5 m_e$ and reanalyzed our data, as shown in Fig. S17. All parameters ($H_i$, $H_{so}$, $A$, $C$) show the same magnitude and trend as the results obtained by assuming $m^* = m_e$; the consistency is particularly good for $V_g < 0$. Therefore, we propose that our fits do not strongly depend on the value of $m^*$, and the variation of $m^*$ within the reasonable range (~0.5-1.0 $m_e$) does not change the qualitative conclusions.

## DATA AVAILABILITY

The data that support the findings of this study are available from the corresponding authors upon reasonable request.

**ACKNOWLEDGEMENTS**

We thank Tao Wu, Zhengyu Wang, Jianjun Ying, and Wei Hu for valuable discussions. This work was supported by the National Key Research and Development Program of the Ministry of Science and Technology of China (2017YFA0303001 and 2019YFA0704901), the National Natural Science Foundation of China (11888101), Anhui Initiative in Quantum Information Technologies (AHY160000), the Science Challenge Project of China (Grant No. TZ2016004), the Key Research Program of Frontier Sciences, CAS, China (QYZDYSSW-SLH021), the Strategic Priority Research Program of Chinese Academy of Sciences (XDB25000000).

**AUTHOR CONTRIBUTIONS**

X. H. and X. C. conceived the experiments; X. H., F. M., Z. L. and Z. H. prepared the interface samples and fabricated the devices; X. H. and Z. H. performed XRD and AFM measurements; X. H. performed the electrical transport measurements; S. W. and B. G. performed the STEM experiments; X. H., F. M., Z. X. and X. C. analyzed the data; X. H., Z. X. and X. C. prepared the manuscript. All authors contribute to editing the manuscript.

**COMPETING INTERESTS**

The authors declare no competing interests.

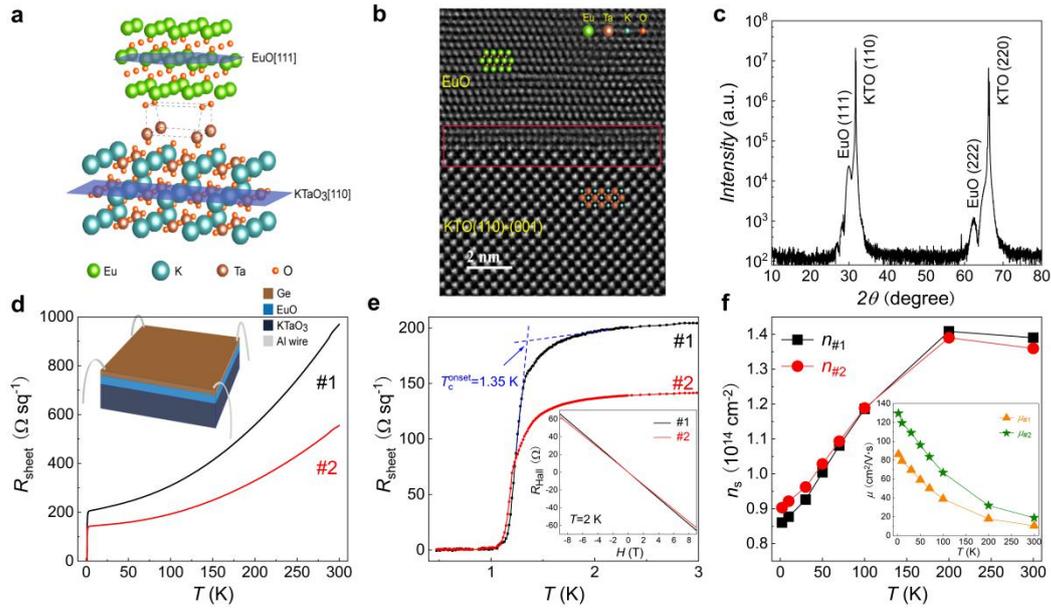

**Fig. 1** **Structural characterization and transport properties of EuO/KTO(110).**
**a** A schematic diagram for the epitaxial growth of EuO (111) on KTO (110). **b** STEM image of the EuO/KTO(110) interface along (001) direction. The red square indicates the interface region. **c** $\theta$-$2\theta$ X-ray diffraction (XRD) pattern specifying the well-orientated EuO (111) film on KTO (110). **d** Sheet resistance as a function of temperature of sample #1 and #2 measured in a wide temperature range. Inset: a sketch of van der Pauw method for the measurements of sheet resistance and Hall resistance. **e** Sheet resistance as a function of temperature shows superconducting transitions at low temperatures. Inset: Hall resistance as a function of magnetic field measured at 2 K. **f** Carrier density $n_s$ and Hall mobility $\mu$ (inset) plotted against $T$.

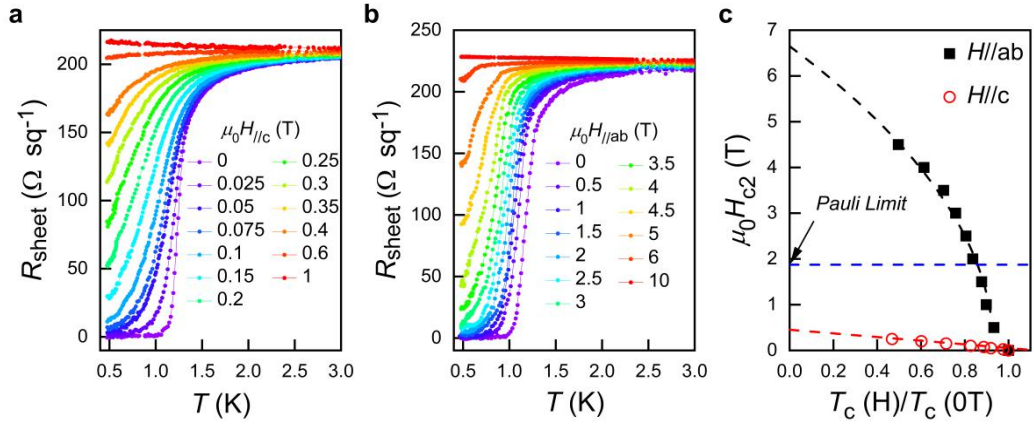

**Fig. 2 Superconducting transition and the upper critical field of EuO/KTO(110).** Temperature-dependent sheet resistance under different magnetic field **a** out-of-plane ($H//c$) and **b** in-plane ($H//ab$). **c** $T/T_c$-dependent upper critical field $\mu_0 H_{c2}$, extracted from the 50% normal-state resistance out-of-plane and in-plane. The estimated Pauli paramagnetic limit ($\mu_0 H_P$) is marked with a blue dashed line in **c**.

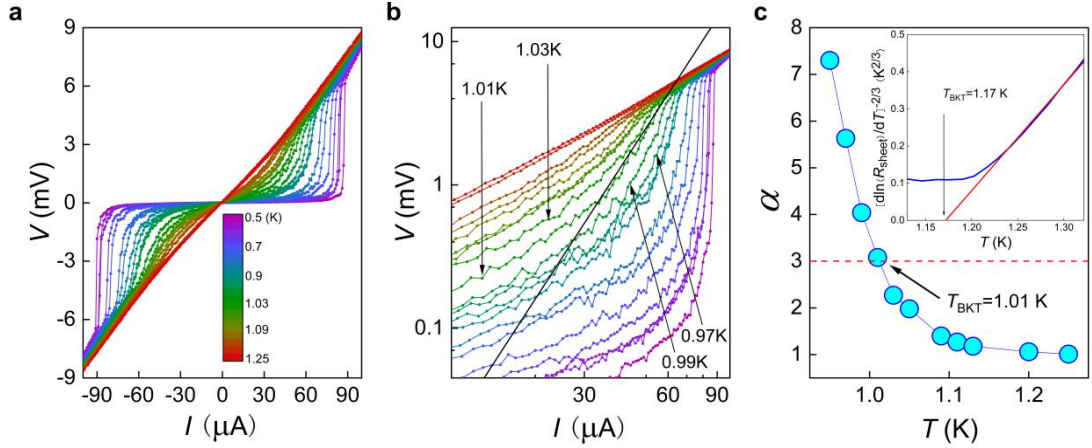

**Fig. 3 The Berezinskii-Kosterlitz-Thouless transition**. **a** *I-V* curves measured at different temperatures. **b** *I-V* curves plotted in a logarithmic-logarithmic scale with the same color codes as in (a). The black solid line represents $V \propto I^3$, which is used to infer the BKT transition temperature $T_{BKT}$. **c** *T*-dependence of the power-law exponent α ($V \propto I^\alpha$) obtained from the linear fits of the curves in **b**, in the range of transition (where the I-V relation is no longer linear). Inset: $[d\ln(R_{sheet})/dT]^{-2/3}$ plotted against *T* [we use the zero-field data of $R_{sheet}(T)$ shown in Fig. 2b]. A linear extrapolation from the high-*T* linear section (red dashed line) crosses the *T*-axis at $T_{BKT}$ = 1.17 K.

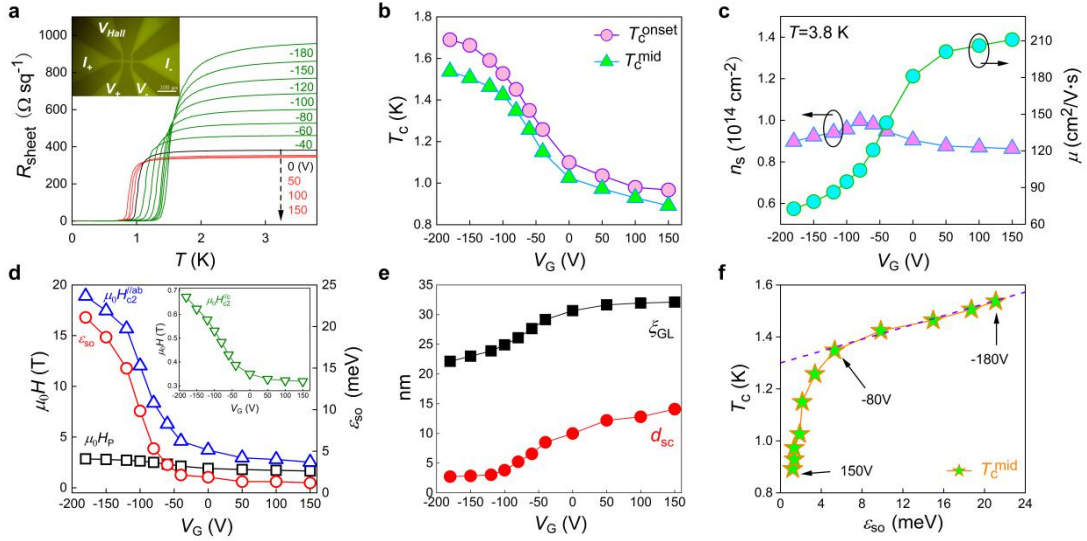

**Fig. 4 Electric field tunability of the superconducting state. a** Temperature-dependent sheet resistance at different gating voltage ($V_G$). Inset: photograph of the device used in the gating process. It was fabricated into a six-probe Hall bar configuration. **b** $V_G$-dependent onset and midpoint of $T_c$. **c** $V_G$-dependent carrier density $n_s$ and Hall mobility $\mu$ measured at $T = 3.8$ K. **d** $V_G$-dependent Pauli limit ($\mu_0 H_P$), in-plane upper critical field ($\mu_0 H_{c2}^{//ab}$), and spin-orbit energy ($\varepsilon_{so}$). Inset: $V_G$-dependent out-of-plane upper critical field ($\mu_0 H_{c2}^{//c}$). **e** $V_G$-dependent GL coherence length ($\xi_{GL}$) and superconducting layer thickness ($d_{sc}$). **f** The spin-orbit energy-dependent midpoint of $T_c$ ($T_c^{mid}$). Purple dashed line is a linear fit to the data at $V_G < -80$ V.

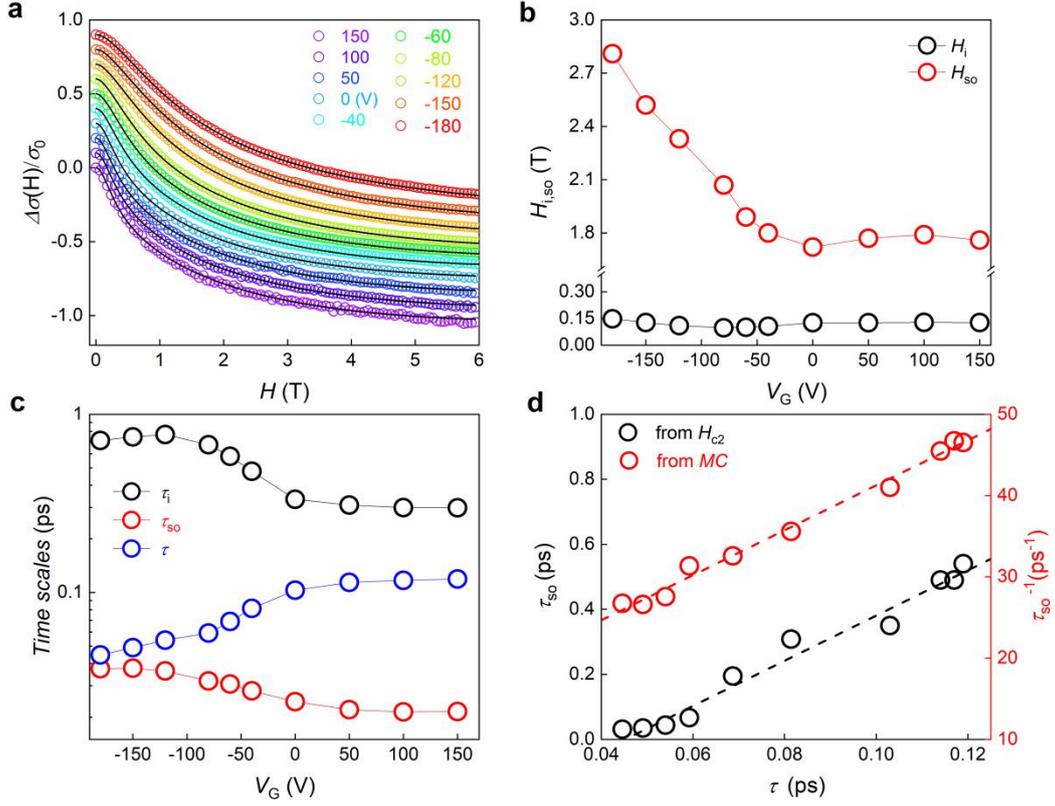

**Fig. 5 Spin-orbit scattering effect in magnetotransport a** The normalized transverse magnetoconductance [$\Delta\sigma(H) = 1/R_{sheet}(H)-1/R_{sheet}(0)$] measured at $T = 3.8$ K under different $V_G$. $H$ is applied perpendicular to interface. We ignore the Hall term due to its small amplitude (Fig. S8). The curves are shifted vertically for clearance. The black solid lines are fits to the Maekawa-Fukuyama model (see text). **b** The $V_G$-dependent effective fields $H_i$ and $H_{SO}$ (see text) extracted from the fitting in **a**. **c** The evolution of the relaxation times for inelastic scattering ($\tau_i$), spin-orbit scattering ($\tau_{so}$), and elastic scattering ($\tau$) upon varying $V_G$. **d** $\tau$-dependent $\tau_{so}^{-1}$ determined from the magnetoconductance (*MC*) and $\tau_{so}$ determined from the upper critical field. The dashed lines are linear fits to guide the eyes.